\documentclass[3p, times,merge]{elsarticle}
\journal{Computer Physics Communications}

\usepackage[plain]{algorithm}
\usepackage{amsmath}
\usepackage{amssymb}
\usepackage{bm}
\usepackage[nolabel]{showlabels}
\usepackage[ISO=false]{diffcoeff}
\usepackage{listings}
\usepackage{graphicx}
\usepackage[utf8]{inputenc}
\usepackage{tablefootnote}
\usepackage{threeparttable}
\usepackage{makecell}
\usepackage{xcolor}
\usepackage{textcomp}
\definecolor{mydarkgray}{gray}{0.2}
\definecolor{darkgreen}{rgb}{0,.5,.0}
\usepackage{tikz}
\usetikzlibrary{fit}
\usepackage{xifthen}

\usepackage{geometry}
\usepackage{subcaption}
\usepackage{hyperref}
\usepackage{caption}
\usepackage{subcaption}
\usepackage{booktabs}

\usepackage{diffcoeff}

\renewcommand{\b}[1]{\mathbf{#1}}

\newcommand{\qt}{{q_{\theta}}}

\newcommand{\eval}[1]{\the\numexpr#1\relax}

\newcommand{\atant}{\operatorname{atan2}}

\lstdefinelanguage{MyPython}{
    keywords={def, for, int, float, return, if, else, range, product}, 
    keywordstyle=\color{red}\bfseries,
    comment=[l]{\#}, 
    string=[b]"
}

\newcommand{\bs}{\mathbf{s}}
\newcommand{\bS}{\mathbf{S}}

\begin{document}

\begin{frontmatter}

\title{Sampling the XY model with an attention-based neural network}
\author[iis]{Piotr Białas}
\ead{piotr.bialas@uj.edu.pl}
\author[kch]{Christian D. Morales}
\ead{m.c.demetrio@student.karazin.ua}
\author[ift]{Tomasz Stebel}
\ead{tomasz.stebel@uj.edu.pl}
\author[iis,ds]{Dawid Zapolski}
\ead{dawid.zapolski@doctoral.uj.edu.pl}

\address[iis]{Institute of Applied Computer Science, Jagiellonian University, ul.~Łojasiewicza 11, 30-348 Kraków, Poland}

\address[kch]{V. N. Karazin Kharkiv National University, 6 Nezalezhnosti Ave, Kharkiv 61022, Ukraine}

\address[ift]{Institute of Theoretical Physics, Jagiellonian University, ul.~Łojasiewicza 11, 30-348 Kraków, Poland}

\address[ds]{Doctoral School of Exact and Natural Sciences, Jagiellonian University, ul.~Łojasiewicza 11, 30-348 Kraków, Poland}

\begin{abstract}
We present a transformer-based variational autoregressive network for
sampling lattice systems with continuous degrees of freedom and apply it to the
two-dimensional XY model near the Berezinskii--Kosterlitz--Thouless transition.
The architecture is combined with physics-informed ingredients: locality-informed attention and approximate probabilities. The resulting sampler significantly outperforms existing neural samplers. We
train the model on lattices up to $96\times96$, verify with a mode-dropping
estimator that no parts of the configuration space are missed, and reproduce
the energy, magnetization and helicity modulus obtained with the traditional Markov Chain Monte Carlo algorithm.
\end{abstract}


\end{frontmatter}

\section{Introduction}
\label{sect_intr}

Markov Chain Monte Carlo (MCMC) remains the method of choice for computing observables in statistical physics and lattice
field theory. Its efficiency, however, is limited by autocorrelations between
consecutive configurations, which grow rapidly in the vicinity of phase
transitions (critical slowing down). For some systems this problem can be
alleviated by cluster algorithms~\cite{PhysRevLett.58.86,PhysRevLett.62.361},
but such algorithms are model-specific and are not available for many systems of
interest, e.g.,\ frustrated magnets, spin glasses or gauge theories. In addition,
MCMC does not give direct access to the partition function and hence to the free
energy and entropy, which require dedicated techniques such as thermodynamic
integration or the Wang--Landau algorithm \cite{PhysRevLett.86.2050}.

In recent years, generative neural networks have been proposed as an
alternative approach. A \emph{neural sampler} is a model that provides both
independent configurations and the exact, normalized probability $q_\theta$ with
which each of them was generated. The network is trained to approximate the
Boltzmann distribution $p$, typically by minimizing the reverse Kullback--Leibler
divergence, which does not require any previously generated training data.
Knowledge of $q_\theta$ allows one to remove the remaining bias between
$q_\theta$ and $p$, either by Neural Importance Sampling (NIS) or by using the
network as an independent proposal in the Metropolis--Hastings algorithm
(Neural Markov Chain Monte Carlo, NMCMC)~\cite{2020PhRvE.101b3304N}. Both
approaches are asymptotically exact, and also  give a direct estimate of the
free energy. Two families of architectures dominate this field: Variational
Autoregressive Networks (VAN)~\cite{2019PhRvL.122h0602W}, which factorize the
probability into a product of conditional distributions of consecutive spins,
and normalizing flows (NF)~\cite{PhysRevD.100.034515}, 
which transform a simple prior distribution by a sequence of invertible maps.

For discrete spin systems, autoregressive networks have advanced considerably. Hierarchical Autoregressive Networks (HAN)~\cite{Bialas:2022qbs,
Bialas:2025hxu} and RiGCS \cite{Singha:2025lsd} exploit the locality of interactions to reduce the number of
spins on which each conditional distribution depends, physics-informed
approximate probabilities~\cite{Biazzo_2023,priors} inject an
analytical approximation of the conditionals directly into the network, and
transformer-based architectures~\cite{2025arXiv250923043B,Bialas:2026imq,scalable} replace
convolutional or dense layers with the attention mechanism~\cite{NIPS2017_3f5ee243}. Systems with continuous degrees of freedom are usually treated with normalizing flows, but the focus has mostly been on lattice field theories \cite{PhysRevD.100.034515, PhysRevLett.126.032001, Gerdes:2022eve, Bialas:2023fyj, Cranmer:2023xbe, Singha:2026slc}. 

The two-dimensional XY model is the natural benchmark for samplers of systems with continuous degrees of freedom: it is the
simplest model with a continuous $O(2)$ symmetry, it exhibits the
Berezinskii--Kosterlitz--Thouless (BKT) transition~\cite{berezinskii,kosterlitz}
driven by the unbinding of vortices, and it shares its compact, $U(1)$-valued
variables with lattice gauge theories. Existing neural samplers for this model
include: Continuous-mixture Autoregressive Networks (CAN)~\cite{Wang_2022},
normalizing flows~\cite{neumc}, and flows trained with
adversarial learning (AdvNF)~\cite{advnf}. As we show in this work,
their sampling quality degrades quickly with the system
size.

In this paper, we follow an approach similar to CAN and sample the XY model with an autoregressive architecture, spin by spin. The probability distribution is parameterized by a linear combination of simple distributions that are easy to sample from. The parameters of this combination are given by the output of the neural architecture. Two crucial differences between our approach and CAN are: (i) instead of a convolutional network, we use a transformer architecture; and (ii) we use von Mises functions that are more physically justified than the beta distributions used in~\cite{Wang_2022}. We combine this
architecture with three physics-motivated ingredients: \emph{locality-informed
attention} \cite{scalable}, which restricts the attention span according to the geometry of the
lattice and the Hammersley--Clifford theorem \cite{ Clifford90markovrandom}; \emph{approximate probabilities} \cite{priors} in the
lowest ($t^1$) order, which for the XY model stay within the von Mises family
and can therefore be combined analytically with the network output; and \emph{averaging the model probability} over a group of lattice symmetries, applied after training.

The paper is organized as follows. In Section~\ref{sect_NS}, we review neural
samplers, asymptotically unbiased estimators, and diagnostics of mode
collapse, and we describe the XY model and the tVAN architecture together with
its physics-informed modifications. Section~\ref{section_results} presents the
numerical results: a comparison with other neural samplers, the effect of
approximate probabilities and gauge fixing, scaling with the lattice size,
physical observables, post-training symmetrization, and tVAN as a
normalizing flow prior. We summarize in Section~\ref{sect_summary}. Derivations
of the $t^1$ approximate probabilities, the
proof that symmetrization does not decrease the ESS, and a study of hyperparameters are collected in the Appendices.

\section{Method}
\label{sect_NS}

\subsection{Neural Samplers}

Let us consider system of $N$ particles, with internal continuous degrees of freedom $s_i$. The probability of finding a system in state $\textbf{s} = \left(s_1,s_2,\ldots,s_N\right)$ is given by the Boltzmann distribution:
\begin{equation}
p(\textbf{s})
=
\frac{1}{Z}
e^{-\beta E(\textbf{s})},
\
\textrm{where}
\
Z
=
\int
d \textbf{s}\,
e^{-\beta E(\textbf{s})}.
\end{equation}
In the above expressions, $E(\textbf{s})$ is the energy of a configuration, $\beta$ is the
inverse temperature, and the integration $\int
d \textbf{s}$ is performed over the whole configuration space (in the case of discrete degrees of freedom, the integral is replaced by a sum).

The main idea of a neural sampler is to generate configurations from some distribution $q_\theta(\textbf{s})$ that approximates well the (target) distribution $p(\textbf{s})$. In practice, the trained distribution $q_\theta(\textbf{s})$ will not reproduce $p(\textbf{s})$ exactly, which leads to a bias in observables. The availability of $q_\theta(\textbf{s})$ allows one to use importance sampling or the Metropolis-Hastings algorithm to correct the remaining bias between $q_\theta(\textbf{s})$ and $p(\textbf{s})$. For this reason, the chosen architecture in this work provides not only configurations but also their normalized probability density.

The free energy can be calculated from the partition function:
\begin{equation}
F=-\frac{1}{\beta}\log Z.
\end{equation}
The model is trained by minimizing the reverse Kullback-Leibler divergence:
\begin{equation}
D_{\mathrm{KL}}(q_\theta\|p)
=
\int
d \textbf{s}\,
q_\theta(\textbf{s})
\log
\frac{q_\theta(\textbf{s})}
     {p(\textbf{s})}\equiv\left\langle
\log \frac{q_\theta(\textbf{s})}{p(\b s)}
\right\rangle_{q_\theta}.
\end{equation}
Substituting the Boltzmann distribution, we get:
\begin{equation}
D_{\mathrm{KL}}(q_\theta\|p)
=
\beta\left(F_q-F\right),
\end{equation}
where
\begin{equation}
\label{F_q_def}
F_q
=
\frac{1}{\beta}
\left\langle
\log q_\theta(\textbf{s})
+
\beta E(\textbf{s})
\right\rangle_{q_\theta}
\end{equation}
is the variational free energy. Since $F$ is independent of $\theta$, minimizing $F_q$ is in fact minimizing $D_{KL}$. Since $D_{\mathrm{KL}}(q_\theta\|p)\geq 0$ it means that $F_q \geq F$.

Minimization of $F_q$ can be done through the REINFORCE algorithm\cite{2019PhRvL.122h0602W}. Let us define the signal:
\begin{equation}
    S_\theta(\textbf{s}) = \log q_\theta(\textbf{s}) + \beta E(\textbf{s}),
\end{equation}
then the gradient of the loss function is written as:
\begin{equation}
\label{gradient_estimator}
    \nabla_\theta F_q = \frac{1}{\beta} \left\langle \left( S_\theta(\textbf{s}) - \left\langle S_\theta (\textbf{s})\right\rangle_{q_\theta}   \right) \nabla_\theta \log q_\theta(\textbf{s}) \right\rangle_{q_\theta}.
\end{equation}

At each optimizer step, a batch of configurations of size $2^{12}=4096$, unless stated otherwise, is generated from $q_\theta$. Energy and log-probabilities of each configuration are then used to calculate the gradient of $F_q$, after which the model's weights are updated.

\subsection{Autoregressive representation}

Let us consider a system of N continuous spins:

\begin{equation}
    \textbf{s} = \left(s_1,s_2,\ldots,s_N\right).
\end{equation}
We can define the joint probability density by factorizing it into a product of conditional densities:
\begin{equation}
    q_\theta\left(s_1, s_2, \ldots, s_N \right) = q\left(s_1\right)\prod_{i=2}^Nq_\theta\left(s_i \mid s_{<i}\right),
\label{eq:probability_density}    
\end{equation}
where
\begin{equation}
    s_{<i} \equiv \left(s_1, s_2, \ldots, s_{i-1} \right).
\end{equation}

The selected model needs to parametrize the conditional densities $q_\theta\left(s_i \mid s_{<i} \right)$, hence it needs to satisfy the autoregressive property - output $i$ may depend only on inputs $j$ that satisfy $j < i$. Configurations are produced using ancestral sampling. 
First, spin $s_1$ is drawn from a uniform distribution on the circle. Then the procedure is iterative: after $s_1, \ldots, s_{i-1}$ are generated, the model evaluates $q_\theta\left(s_i \mid s_{<i}\right)$ and draws the next spin from this density. The process continues until all of the $N$ spins are generated.

\subsection{Asymptotically unbiased observables}

The VAN algorithm described above does not guarantee correct sampling from $p$. It only provides a \emph{variational upper bound} on
thermodynamic quantities such as the free energy. For this reason, the variational estimate
alone is never sufficient, and it must be supplemented by one of the
asymptotically exact procedures \cite{2020PhRvE.101b3304N}: Neural Markov Chain
Monte Carlo (NMCMC) or Neural Importance Sampling (NIS). 

 NMCMC introduces an accept/reject step: a configuration proposed by the network, $\mathbf{s}_{\textrm{next}}$, is accepted with probability
\begin{equation}
\label{eq prob}
p_{acc}=\min\left( 1, \frac{q_{\theta}(\mathbf{s}_{\textrm{previous}})
p(\mathbf{s}_{\textrm{next}})}{q_{\theta}(\mathbf{s}_{\textrm{next}})
p(\mathbf{s}_{\textrm{previous}})} \right),
\end{equation}
which is the Metropolis--Hastings rule for an independent proposal
distribution.
 
The NIS compensates for the mismatch between $p$ and $q_\theta$ by assigning each configuration an importance weight:
\begin{equation}
\label{eq. weight}
w(\mathbf{s}) = \frac{e^{-\beta E(\mathbf{s})}}{q_\theta(\mathbf{s})} \,.
\end{equation}

The expectation value of an observable $\mathcal{O}$ in the target
distribution can then be rewritten as an expectation value in the
distribution generated by the network,
\begin{equation}
\label{eq. reweighting}
\langle \mathcal{O} \rangle_p = \int d\mathbf{s}\, p(\mathbf{s})\,
\mathcal{O}(\mathbf{s})
= \frac{\int d\mathbf{s}\, q_\theta(\mathbf{s})\, w(\mathbf{s})\,
\mathcal{O}(\mathbf{s})}{\int d\mathbf{s}\, q_\theta(\mathbf{s})\,
w(\mathbf{s})} \,.
\end{equation}
In practice, integrals over the whole configuration space are estimated on a set of $N_{stat}$ configurations
$\mathbf{s}_i$ drawn independently from the distribution $q_\theta(\bs)$:
\begin{equation}
\label{eq. nis estimator}
\langle \mathcal{O} \rangle_p \approx
\frac{\sum_{i=1}^{N_{stat}} w(\mathbf{s}_i)\,
\mathcal{O}(\mathbf{s}_i)}{\sum_{i=1}^{N_{stat}} w(\mathbf{s}_i)}, \quad \mathbf{s_i}\sim q_\theta.
\end{equation}
When $N_{stat}\rightarrow\infty$ this estimator converges to the correct value for any $q_\theta$, provided
only that $q_\theta(\mathbf{s})>0$ wherever $p(\mathbf{s})>0$, so that no
region of the configuration space is inaccessible to the network. In this
sense NIS, exactly like NMCMC, turns an imperfectly trained network into an
asymptotically exact algorithm; the quality of the training determines the
statistical error, not the correctness of the result.

The weights give direct access to the
partition function itself,
\begin{equation}
\label{eq. Z estimate}
Z_{NIS} = \int\!\dl\bs\,q_\theta(\mathbf{s})\, w(\mathbf{s}) \approx
\frac{1}{N_{stat}} \sum_{i=1}^{N_{stat}} w(\mathbf{s}_i),\quad \textbf{s}_i\sim q_\theta(\bs) \,,
\end{equation}
and hence to the free energy 
\begin{equation}
\label{F_nis_eq}
F_{NIS} = -\frac{1}{\beta}\log Z_{NIS}. 
\end{equation}

The difference with
respect to the variational estimate $F_q$ of Eq.~\eqref{F_q_def} is essential.
The estimate \eqref{eq. Z estimate} is unbiased, and although taking its
logarithm introduces a bias, it is of order $1/N_{stat}$ and can be
made arbitrarily small by increasing the statistics. The gap between $F_q$ and
$F$, on the contrary, is a property of the trained network and does not
decrease with the number of generated configurations.

The price to be paid is that the variance of the weights
\eqref{eq. weight} may be very large. If $q_\theta$ differs appreciably from
$p$, a few configurations with huge weights dominate the sums in
Eq.~\eqref{eq. nis estimator} and the effective statistics is much smaller
than $N_{stat}$. This is quantified by the effective sample size,
\begin{equation}
\label{eq. ess}
\textrm{ESS} = \frac{\left(\sum_{i=1}^{N_{stat}} w(\mathbf{s}_i)\right)^2}
{N_{stat}\,\sum_{i=1}^{N_{stat}} w(\mathbf{s}_i)^2} \in (0,1] \,,
\end{equation}
which equals one when $q_\theta = p$ and tends to zero when the network
distribution and the Boltzmann distribution differ strongly. The ESS is
therefore both a measure of the quality of the trained model and the quantity
which has to be monitored and reported together with every result: it plays,
for NIS, the role that the inverse integrated autocorrelation time plays for NMCMC. 

It should be
stressed that a large ESS is a necessary but insufficient condition for
correctness. If the network never generates configurations belonging to part of the configuration space, the weights cannot repair this, since
reweighting can only redistribute the probability among configurations that
have actually been generated. The same is true for the NMCMC method. Such mode collapse is particularly dangerous in
systems where the structure of modes is not well known, for example, spin glasses. In practice, even if formally all configurations have $q_\theta>0$, the mode collapse can still occur with finite $N_{stat}$.

In general, there is no method to detect mode collapse other than by comparing results with other methods. Assuming that one has access to samples generated from distribution $p$, it is possible to measure mode collapse. In Ref. \cite{DetectingModeColapse}, it was proposed to use the following estimate of free energy in comparison to (\ref{F_nis_eq}):
\begin{equation}
\label{FMC_def}
    F_{MC} = -\frac{1}{\beta}\log\left(Z_{MC}\right)
\end{equation}
where 
\begin{equation}
    Z^{-1}_{MC} = \left\langle \frac{q_\theta(\mathbf{s})}{e^{-\beta E(\mathbf{s})}} \right\rangle _p
    \approx 
    \frac{1}{N_{stat}}\sum_{i=1}^{N_{stat}}\frac{q_\theta\left(\textbf{s}_i\right)}{e^{-\beta E(\mathbf{s}_i)}}, 
    \quad
    \textbf{s}_i \sim p
\end{equation}
It can be shown that \cite{DetectingModeColapse}:
\begin{equation}
F_{NIS}\ge F \ge F_{MC},
\end{equation} 
hence the difference between $F_{NIS}$ and $F_{MC}$ is a measure of mode seeking. Following \cite{DetectingModeColapse}, we define the mode-dropping estimator:
\begin{equation}
\label{mode_dropping_estim}
\mathcal{W} \equiv \frac{Z_{NIS}}{Z_{MC}} = e^{\beta(F_{MC}-F_{NIS})}. 
\end{equation}
A value of $\mathcal{W}$ below one indicates missing modes.

In this contribution, we shall use samples generated using an in-house Wolff MCMC code which is an implementation of \cite{PhysRevLett.62.361}. This script performs $10^4$ thermalization cluster updates and stores $10^6$ configurations that are separated by $n_{sep}$ cluster updates. The $n_{sep}$ was chosen so that consecutive stored samples have $\tau_{int} \leq 1.02$.
This means  $n_{sep} =70, 80, 90$ for $L= 16, 32, 64$; $n_{sep} =90$ for  $L=96$ at $\beta = 1.0$ or $1.12$; $n_{sep}=190$ for $L=96$ at $\beta = 1.2$. 
Our script was checked against \cite{Hasenbusch_2005} at $\beta = 1.1199$ and lattice sizes up to $L=128$ and the observables agree within statistical error.

\subsection{XY model}

In this work we consider a 2D XY model defined on an $L\times L$ lattice, implying $N=L^2$. Each lattice site contains a two dimensional spin vector $\bS_i$ of length one, and the 
energy of the configuration is defined as
\begin{equation}
    E(s) = -J \sum_{\left\langle i,j \right\rangle}\vec{S_i}\cdot \vec{S_j},
\label{eq:xy_energy}
\end{equation}
where only the nearest-neighbor interactions are included.
Spins at each lattice site can be  represented as 
two-component vectors\footnote{In what follows, we refer to $s_i$ as spins for simplicity, although they formally represent angular variables. Since there is a one-to-one correspondence between $s_i$ and the spin vectors $\vec S_i$, this terminology should not cause any ambiguity.}
\begin{equation}
    \vec{S}_i
    =
    \left(\cos s_i,\sin s_i\right),
    \qquad
    s_i\in [0,2\pi).
    \label{eq:xy_spin}
\end{equation}
We can rewrite Eq.~\eqref{eq:xy_energy} as 
\begin{equation}
    E(s) = - J \sum_{\left\langle i,j \right\rangle}\cos(s_i - s_j).
\end{equation}
In this work, we assume that the couplings are constant $J=1$, and the periodic boundary conditions are imposed. In the thermodynamic limit, the model exhibits the famous BKT transition~\cite{berezinskii,kosterlitz} around an inverse temperature of $\beta = 1.12$.

Since the energy depends only on the angle differences, the model itself is invariant under global rotation by $\alpha$, meaning
\begin{equation}
    s_i\longrightarrow (s_i+\alpha)\bmod 2\pi,
    \qquad
    E(s_1+\alpha,\ldots,s_N+\alpha)
    =
    E(s_1,\ldots,s_N).
    \label{eq:global_rotation}
\end{equation}
This is a global gauge symmetry, and we can use it to fix the first spin to $s_1 = 0$ during generation. 
The contribution of $q(s_1)$ to the probability density $q_\theta$ is retained and equals that of a uniformly drawn spin:
\begin{equation}
    q(s_1) = \frac{1}{2\pi}
\end{equation}
This allows us to rewrite the logarithm of Eq.~\eqref{eq:probability_density} as
\begin{equation}
    \log q_\theta(\mathbf{s}) = - \log 2\pi + \sum_{i=2}^N \log q_\theta\left(s_i \mid s_{<i}\right)
\end{equation}

\subsection{Transformer architecture}

An autoregressive transformer \cite{NIPS2017_3f5ee243} with causally masked attention outputs a sequence of elements by predicting each of them from the previous ones. Usually, in this type of model, elements are treated as tokens from a finite vocabulary, and the output at a given position is a categorical distribution over the values from said vocabulary. The tVAN \cite{Bialas:2026imq} that was previously introduced for discrete spin systems, specifically the Ising and Edwards--Anderson models, uses this exact idea: the spins or patches of spins are treated as tokens in the model's vocabulary, and the network assigns a probability to each existing token. 

This approach cannot be used for an XY model directly without binning the spins, since each lattice site carries a continuous value $s_i\in[0,2\pi)$ and every $q_\theta\left(s_i \mid s_{<i} \right)$ is a probability density rather than a finite set of probabilities. 
In order to utilize tVAN's approach, we need to adapt the interface through which the model ingests spin values and outputs spin probability densities. 

At the input, the spin needs to be represented by some form of a continuous value. A simple scalar representation of the angle introduces a continuity problem at the $0 \xrightarrow{} 2\pi$; two physically close spins would have numerically distant representations. Therefore, every spin is represented as a pair 
$ \left(\cos s_i, \sin s_i \right)$.
The two components are mapped to the embedding space, and a learned positional embedding is used to provide the network location of spins on different lattice sites.

The embedded sequence is processed by transformer blocks. Each block consists of multi-head self-attention, a position-wise feed-forward network, and residual connections with layer normalization. A causal attention mask ensures that the network uses only previously generated spins to construct $q_\theta\left(s_i\mid s_{<i}\right)$. Attention used in this work additionally restricts the context by following the geometry of the lattice and will be described in the following section.

At the output, the transformer's result determines the conditional density $q_\theta\left(s_i\mid s_{<i}\right)$. In this work we parametrize it with a mixture of von Mises distributions,
\begin{equation}
    \label{param_q_vonMises}
    q_\theta\left(s_i\mid s_{<i}\right) = \sum_{j=1}^K \omega_{ij} 
     \operatorname{VM}\left(s_i; \mu_{ij}, \kappa_{ij} \right),
\end{equation}
where
\begin{equation}
    \operatorname{VM}\left(s; \mu, \kappa \right) = \frac{e^{\kappa \cos\left(s-\mu\right)}}{2\pi I_0\left(\kappa\right)}, \qquad \kappa \geq 0
\end{equation}
and  $I_0$ is a modified Bessel function of the first kind.

The choice of von Mises distribution is dictated by the fact that it closely resembles the Boltzmann distribution of the energy of the XY model. In principle, any distribution can be used for this purpose, but choosing von Mises allows for efficient $q_\theta$ representation even for small values of $K$. To sample from the von Mises distribution, we utilize inverse-CDF sampling~\cite{devroye1986} instead of a standard rejection algorithm, this gives every sample the same cost, without introducing rejections, and makes sampling more efficient on the GPU.

The transformer produces $3K$ real-number outputs at each site that we will label as $a_{ij}$, $b_{ij}$, $c_{ij}$, which are connected to von Mises parameters     \eqref{param_q_vonMises} by:
\begin{equation}
\label{output_transf}
    \omega_{ij} = \operatorname{softmax}\left(a_i\right)_j, \qquad \mu_{ij} = b_{ij}, \qquad \kappa_{ij} = \operatorname{softplus}\left(c_{ij}\right).
\end{equation}

\subsection{Locality-informed attention}

Causal attention allows the model to generate an output that is dependent only on previously generated tokens. Usually this implies that the model's output uses all of the preceding tokens, but in the case of the XY model, this is not strictly necessary. Since Eq.~\eqref{eq:xy_energy} contains only nearest-neighbor interactions, one can show that restricting the attention span of the model while considering the lattice's geometry does not reduce tVAN's quality. Formally, this is a consequence of the Hammersley-Clifford theorem \cite{Clifford90markovrandom}
and was used to propose Hierarchical Autoregressive Networks (HAN) \cite{Bialas:2022qbs, Bialas:2025hxu}.
In the context of the Ising model and transformer-based architectures, it was first used in \cite{scalable}.

\begin{figure}
    \begin{center}
\begin{tikzpicture}[scale=0.9,thick , x=1cm, y=1cm, every node/.style={circle, draw, minimum size=7mm, inner sep=0pt, scale=1.0, font=\small}]
  \def\L{6}
  \def\rows{\L}
  \def\cols{\L}
  \def\offset{\L}
\definecolor{lightgray}{rgb}{0.9,0.9,0.9}
  \foreach \i in {0,...,\eval{\rows-1}} {
    \foreach \j in {0,...,\eval{\cols-1}} {
      \newcommand{\n}{\eval{\cols*\i+\j+1}}
      \newcommand{\lab}{\n}
     
      \ifthenelse{\n=21}{\node [double, draw=red] (s\n) at (\j, \offset-\i) {$s_{\lab}$};}{
        \ifthenelse{\n<21}
        {
        \ifthenelse{\n<\eval{\cols+1} \OR \n>14} 
        {\node[fill=lightgray] (s\n) at (\j, \offset-\i) {$s_{\lab}$};}
        {\node (s\n) at (\j, \offset-\i) {$s_{\lab}$};
        }
        }
        {\node [dashed] (s\n) at (\j, \offset-\i) {$s_{\lab}$};}
      }
    }
  }

\draw [black, very thick] (s20)--(s21)--(s15);

\end{tikzpicture}
\end{center}
    \caption{Schematic view of spin-by-spin configuration generation, for $L=6$. Solid circles represent already generated spins, and the red spin ($s_{21}$) is about to be generated. Locality-informed attention means that the conditional probability of a spin depends only on two rows of previously generated spins (darker circles). The approximate probabilities depend on two spins: top and left (for spin $s_{21}$ this means $s_{21,top}=s_{15}$  and $s_{21,left}=s_{20}$). 
    } 
    \label{fig:p-cond-27}
\end{figure}
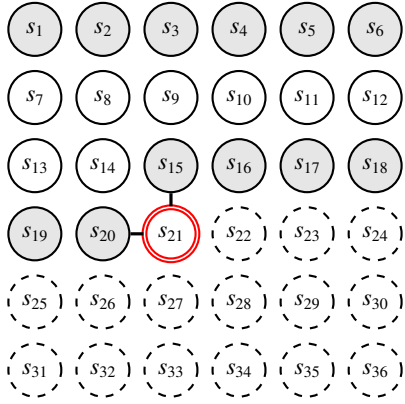

We can rewrite our conditional density as follows
\begin{equation}
    p\left(s_i\mid s_{<i}\right) = p\left(s_i\mid s_{B_i}\right)
\end{equation}
where 
\begin{equation}
    B_i = \{ j < i \mid j \leq L\;\; or\;\; i - j \leq L\},
\end{equation}
 which consists of the most recently generated $L$ spins and the entire first row - see Figure~\ref{fig:p-cond-27} where darker circles represent $B_{21}$ in a $6\times 6$ system.

\subsection{Approximate probabilities}

The approach of using probability priors to inject physics awareness into the model was developed in \cite{priors} (see also \cite{Biazzo_2023} for a similar approach to physically inspired improvements). It introduces a simple approximation to the conditional distribution directly into the neural sampler, which now leaves the network to learn only the remaining difference. 

In this work, we consider priors depending only on two nearest-neighbor spins - this is the lowest-order $t^1$ approximation considered in \cite{priors}. Higher-order approximations were considered, but $t^2$ no longer stays in the von Mises family. The derivation is given in \ref{t1_priors_section}; here we present the final result.

The conditional density with approximate probabilities (AP) can be expressed as
\begin{equation}
\label{eq_q_decomp_AP}
    q_\theta\left(s_i\mid s_{<i}\right) = \sum_{j=1}^K \tilde \omega_{ij}
    \operatorname{VM}\left(s_i; \, \tilde \mu_{ij}, \, \tilde \kappa_{ij}\right),
\end{equation}
where 
\begin{equation}
    \tilde \omega_{ij} = {\displaystyle \omega_{ij} \frac{I_0(\tilde \kappa_{ij})}
    {I_0\left(\kappa_{ij}\right)} }\left(\displaystyle \sum_{l=1}^K \omega_{il}\frac{ I_0(\tilde \kappa_{il})}
    {I_0\left(\kappa_{il}\right)} \right)^{-1},
\end{equation}

\begin{equation}
       \tilde \kappa_{ij} = \sqrt{\tilde A_{ij}^2 + \tilde B_{ij}^2}, \qquad \tilde \mu_{ij}  = 
       \atant\left(\tilde B_{ij},\tilde A_{ij}\right),
\end{equation}

\begin{equation}
\begin{aligned}
    \tilde A_{ij} &= \kappa_{ij}\cos(\mu_{ij}) +\kappa_{i}^{AP}\cos(\mu_i^{AP}), \\
    \tilde B_{ij} &= \kappa_{ij}\sin(\mu_{ij}) +\kappa_{i}^{AP}\sin(\mu_i^{AP}),
\end{aligned}
\end{equation}

\begin{equation}
 \kappa_i^{\mathrm{AP}} = \beta \sqrt{A_i^2 + B_i^2}, \qquad \mu_i^{AP} = 
 \atant\left(B_i,A_i\right).
\end{equation}

In the above expressions $\omega_{ij}$, $ \mu_{ij}$, $\kappa_{ij}$ are obtained from the output of the transformer according to Eq.~\eqref{output_transf}, whereas $A_i$ and $B_i$ are directly computed from two neighboring, already fixed spins
\begin{equation}
    A_i =  \cos(s_{i,\,left}) + \cos(s_{i,\,top}), \qquad B_i= \sin(s_{i,\,left}) + \sin(s_{i,\,top}),
\end{equation}
where we used the notation $s_{i,\,top}$ and $s_{i,\,left}$ since spins are generated in the raster order (see Figure  \ref{fig:p-cond-27}).

\section{Results}
\label{section_results}

All presented results in this section were obtained with the same transformer architecture: a single transformer layer, an embedding dimension of $128$, and $4$ attention heads. We also use $K=4$, which means that for each spin the output of the transformer is 12 real numbers.
For the optimizer, we used AdamW with $\beta_1=0.9$, $\beta_2=0.95$ and a constant learning rate 0.001. In \ref{appendix_hyperparameters} we discuss the dependence of training efficiency on the number of layers and $K$.

\subsection{Comparison with existing neural samplers}
\label{can_nf_tvan}

We start by comparing tVAN with two neural samplers that have already been used for continuous lattice systems: Continuous-mixture Autoregressive Network (CAN) introduced in Ref.~\cite{Wang_2022} and a normalizing-flow sampler~\cite{neumc}. 

CAN was introduced as a model for continuous spin systems and was applied to the two-dimensional XY model, using PixelCNN as its autoregressive representation and utilizing a mixture of Beta distributions to represent the autoregressive conditional. In this comparison, we used the implementation of Ref.~\cite{Wang_2022}: $3$ masked convolutional layers with $32$ channels and kernel of size $3\times3$. The conditional density is constructed as a mixture of $50$ Beta distributions.

An approach using normalizing flows ~\cite{2019arXiv190809257K} represents a probability density through an invertible transformation of some prior distribution. We use the NeuMC package \cite{neumc} to construct a model with $16$ circular-spline\cite{flowsontori} coupling layers. We used splines with nine knots. 
Each layer contains four convolution layers with a $3\times3$ kernel and 16 channels with $\operatorname{LeakyReLU}$ activation functions in between. The prior distribution was uniform on $[0,2\pi)$ for each spin $s$.
We minimized the reverse Kullback–Leibler divergence using the path-gradient estimator \cite{vaitl2022gradients} and optimized the parameters with Adam at a learning rate of $5 \times 10^{-4}$.

\begin{figure}[t]
    \centering
    \includegraphics[width=0.6\linewidth]{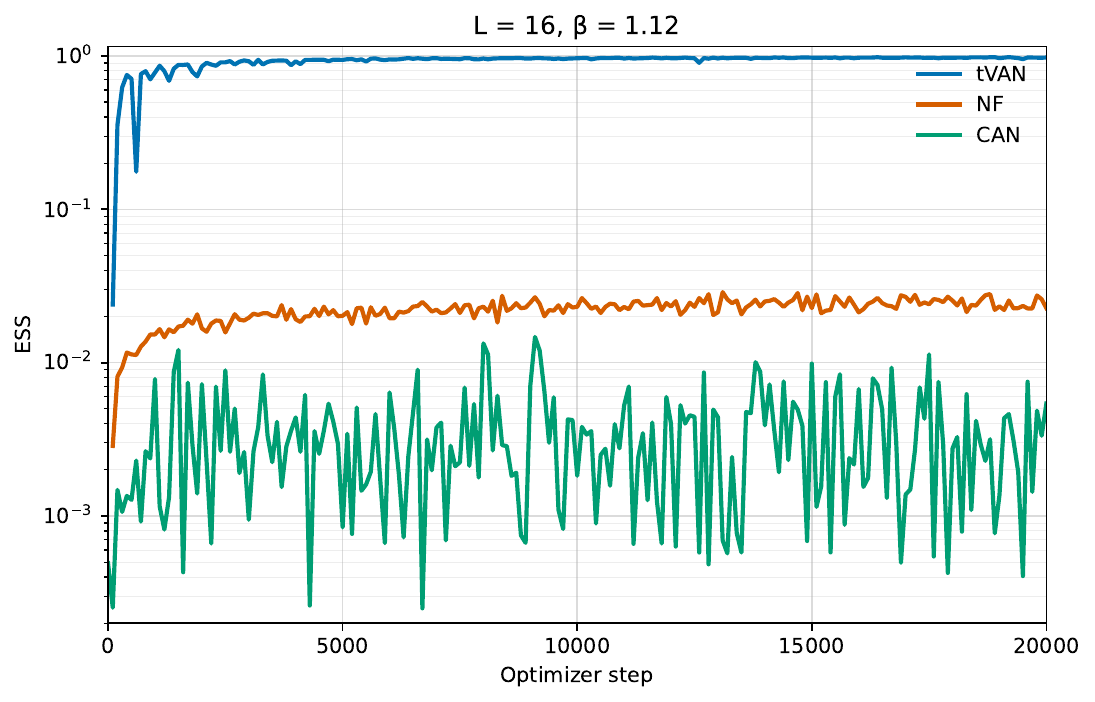}
    \caption{ESS comparison of CAN, NF, and tVAN through optimizer steps.}
    \label{fig:can_nf_tvan_comparison}
\end{figure}

Figure~\ref{fig:can_nf_tvan_comparison} shows a comparison of ESS across all three models at a lattice size of $L=16$ and an inverse temperature of $\beta = 1.12$. The horizontal axis denotes optimizer steps, and the vertical axis represents ESS on a logarithmic scale. The training of the CAN architecture does not increase the ESS significantly, as ESS is volatile    and sits at around $10^{-3}$~--~$10^{-2}$, whereas NF plateaus at $10^{-2}$, and tVAN rapidly approaches $1$ and plateaus. For bigger system sizes, CAN and NF fail to train.

A stronger baseline can be found in the work of \cite{advnf}, which uses Adversarial Learning for Normalizing Flows ($\operatorname{AdvNF}$). This study provides an average acceptance rate (AR) of the Markov chain \eqref{eq prob} as an evaluation metric for the sampler. At size $L=32$, $\operatorname{AdvNF}$ achieves the best AR of $7.2\%$ when averaged over $10$ temperatures $T \in [0.85, 1.25]$. In comparison, tVAN reaches AR of $80.48\%$ at the same $L=32$ directly at $\beta = 1.12$.  

\subsection{Physics-informed improvements}

 In this section, we will compare how injecting probability priors and using gauge fixing affect training efficiency. This comparison consists of four different trained models at $L=32$ and $\beta=1.12$, covering all combinations of gauge fixing and AP switched on and off. For readability, the ESS was computed as a 500-step mean.

\begin{figure}[t]
    \centering
    \includegraphics[width=0.7\linewidth]{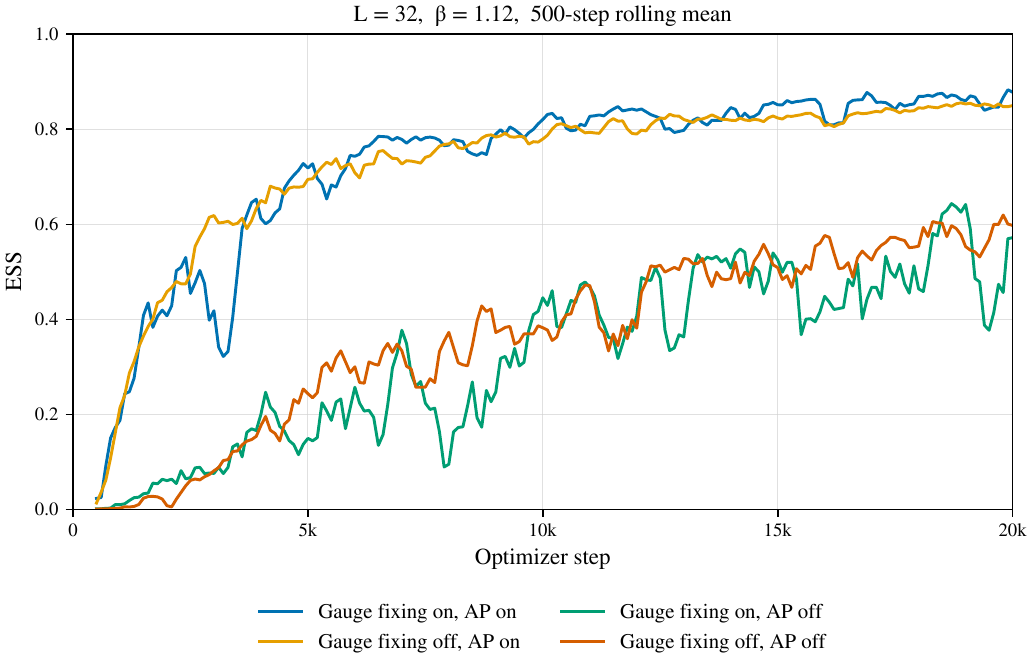}
    \caption{Effect of gauge fixing and AP on ESS through optimizer steps.}
    \label{fig:ess_per_step_rolling_mean_500}
\end{figure}

Figure~\ref{fig:ess_per_step_rolling_mean_500} shows all four of the training curves for each of the model types. From the graph we can clearly see that the AP contribution to training speed and quality is significant. AP increases the ESS by a factor of $1.5$--$1.8$ after $20k$ steps. On the other hand, global gauge-fixing yields no visible ESS advantage.

\subsection{Results for different lattice sizes}

We now explore the results of tVAN on different lattice sizes. The lattice sizes span from $L=16$ to $L=96$. Models are trained at $\beta = 1.12$, utilizing spin-fixing and enabling AP throughout all of the runs. For each lattice size, we trained the model for $2\times 10^4$ steps. After training, $10^6$ configurations are collected from the model and used to measure ESS. To calculate $F_{MC}$ \eqref{FMC_def} we used $10^6$ Wolff configurations.

Table~\ref{tab:perfomance_across_sizes} shows ESS and $\mathcal W$ throughout the lattice sizes. We can see that at $L=16$ tVAN reaches ESS of $0.976$, which decreases to $0.834$ at $L=32$. The degradation becomes considerably stronger as the lattice size increases with the model reaching an ESS of $0.160$ at $L=64$ and $0.074$ at $L=96$. On the other hand, across all of the entries $\mathcal W$ remains consistent with 1, which indicates that this drop in ESS is not related to a problem with missing modes. This is important, as it shows that NIS/NMCMC can compensate for the difference between $q_\theta$ and $p$.

\begin{table}[t]
  \centering
  \begin{tabular}{cccccc}
  \toprule
  $L$ & $\beta$ & ESS
  & $\mathcal{W}$ \\
  \midrule
  16 & 1.12
  & 0.975935
  & $0.99997 \pm 0.00022$ \\

  32 & 1.12
  & 0.833663
  & $1.00029 \pm 0.00062$ \\

  64 & 1.12
  & 0.159995
  & $0.99961 \pm 0.00238$ \\

  96 & 1.12
  & 0.074316
  & $0.99828 \pm 0.00556$ \\

  \bottomrule
  \end{tabular}
    \caption{
  ESS and mode-dropping estimator comparison using $10^6$ tVAN samples and
  $10^6$ Wolff configurations.
  }
  \label{tab:perfomance_across_sizes}
  
  \end{table}

\subsection{Physical observables}

While ESS provides a good metric for estimating the model's training and sampling performance, and $\mathcal W$ estimates the quality of the sampler in terms of mode collapse, we have not yet established if the model is capturing the underlying system's physics. To assess this, we compare the model's predictions with $\operatorname{MCMC}$ on a $96\times96$ lattice at inverse temperatures of $\beta \in \{1.0, 1.12, 1.2\}$. As observables, we compare the mean energy $\langle E \rangle$ and the mean absolute magnetization
\begin{equation}
    |m|=\frac{1}{N} \left\langle \left|\sum_i \textbf{S}_i\right|\right\rangle_p,
\end{equation}
where $N$ is the number of lattice sites.

We also define the helicity modulus $\Upsilon$~\cite{2003PhRvB..67q2509M}
\begin{equation}
    \Upsilon = -\frac{1}{L^2}\left\langle E(\b s)/2 \right\rangle_p - \beta L^2\left\langle \left(\mathcal{S}_x^2(\b s)+ \mathcal{S}_y^2(\b s)\right)/2\right\rangle_p,
    \label{eq_helicity_def}
\end{equation}
where 
\begin{equation}
    \mathcal{S}_x(\b s) = \frac{1}{L^2}\sum_{\left\langle i,j\right\rangle_x} \sin\left( s_i - s_j\right)\quad \textrm{and} \quad \mathcal{S}_y(\b s) = \frac{1}{L^2}\sum_{\left\langle i,j\right\rangle_y} \sin\left( s_i - s_j\right).
\end{equation}
In the above expressions, $\sum_{\left\langle i,j\right\rangle_x}$ ($\sum_{\left\langle i,j\right\rangle_y}$) means that the sum is over all links in the horizontal (vertical) direction. Definition \eqref{eq_helicity_def} is equivalent to the one from Ref.~\cite{2003PhRvB..67q2509M} because our system is symmetric with respect to interchanging the directions $x\leftrightarrow y$. We verified this by explicitly comparing our results for $\Upsilon$ with those from Ref.~\cite{2003PhRvB..67q2509M}.

In addition to physical observables, we also consider the mode-dropping estimator \eqref{mode_dropping_estim}.

\begin{figure}[t]
    \centering
    \includegraphics[width=\linewidth]{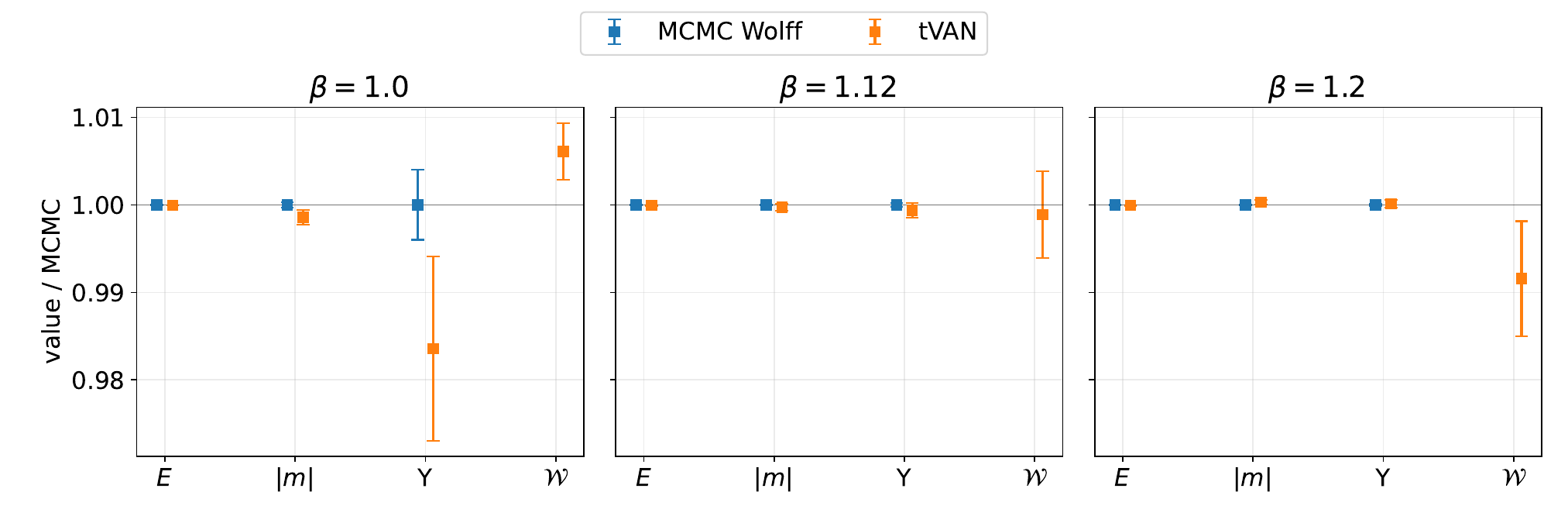}
    \caption{Energy ($E$), absolute magnetization ($|m|$), helicity modulus $\Upsilon$, and $\mathcal{W}$ at lattice size $L=96$ across $\beta \in \{1.0, 1.12, 1.2\}$.}
    \label{fig:l96_energy_mag_helicity_w_across_betas}
\end{figure}

Figure~\ref{fig:l96_energy_mag_helicity_w_across_betas} provides this comparison for three $\beta$ at $L=96$.  All mean values and uncertainties are divided by the MCMC mean, so that all observables can be shown on a single plot. For calculating each observable, we use $10^6$ tVAN and $10^6$ Wolff configurations. Errors were calculated using the bootstrap method.

For all three inverse temperatures $\beta$, we can clearly see that tVAN observables reproduce $\operatorname{MCMC}$ results for all three observables up to statistical fluctuations. 

\subsection{Model symmetrization}
\label{sec:symmetrization}

Up to now, the results we have presented have been obtained using configurations and their probabilities $q_\theta$. The autoregressive factorization explicitly breaks many lattice symmetries that the system possesses due to the periodic boundary conditions. Ref.~\cite{2019PhRvL.122h0602W} suggested, in the context of VAN, using the averaged probability $q_\theta$ during training so that the model is exposed to flipped spin configurations $\mathbf{s}\rightarrow -\mathbf{s}$. This approach can be trivially extended to any system symmetry and has often been used in our previous works~\cite{Bialas:2021bei,Bialas:2022bdl,Bialas:2023fjz}. Here we will simplify the method and apply symmetrization only \textit{after} training. This means that the training is unaltered compared to previous sections; however, at the generation stage, when one assesses the sampler's accuracy or calculates physical observables, the probability density is averaged. In \ref{appendix_symmetriza}, we show a proof that such symmetrization leads to an ESS no smaller than the original one.

For a given configuration $\bf s$, we define the averaged probability: 
\begin{equation}
\label{symmmetrization_q}
    \displaystyle \bar q_\theta (\textbf{s}) = \frac{1}{|G|} \sum_{g \in G} q_\theta(g \textbf{s}),
\end{equation}
where $G$ is a finite (with $|G|$ elements) group of symmetry transformations, i.e. $p(g \textbf{s})=p( \textbf{s}), \, g\in G$. In what follows, we shall use groups of square lattice symmetries with periodic boundary conditions. These include 
the translations $T_n$, where the lattice is shifted by $kL/n, \, k=1,2,\ldots, n-1$ sites in both directions; the rotations of the lattice $C_4$; the reflection of the lattice along the horizontal axis $M_2$; and finally, $D_4$, the dihedral group of order 8 (a group of a square's symmetries). 

It is worth noting that in the VAN approach, in particular the gradient estimator \eqref{gradient_estimator}, 
 the configuration $\textbf{s}$ is first generated from the model spin by spin, and then the probability of this configuration  $q_\theta(\textbf{s})$ is calculated on the whole configuration. The difference between the forward pass through the network for generating configurations and the evaluation of their probabilities lies in the gradient tracing: the former is done without gradients, while the latter needs them, which significantly slows down the pass. For these reasons, using \eqref{symmmetrization_q} during training for large $|G|$ is numerically costly. On the other hand, post-training generation does not require gradient tracing; hence, symmetrization is less expensive.

The symmetrization requires substituting $q_\theta \to \bar q_\theta$ in all definitions we introduced in previous sections, in particular, the definitions of $F_q$ and ESS. The average $\left\langle \ldots \right\rangle _{\bar q_\theta}$ requires sampling from $\bar q_\theta$. This, in practice, can be done by first sampling configurations from $q_\theta$ and then randomly applying a given transformation $g\in G$, with the probability of applying it equal to ${1}/{|G|}$. When the observable $\mathcal{O}$ is invariant under $G$, the procedure of randomly applying group transformations doesn't change the result, hence 
\begin{equation}
\label{q_bar_sampling}
\left\langle \mathcal{O} \right\rangle _{\bar q_\theta}=\left\langle \mathcal{O} \right\rangle _{q_\theta}.  
\end{equation}
In the cases we will consider in this manuscript, this condition is satisfied.\footnote{If the observable depends on the probability distribution, as in the case of $F_q$ or ESS, one needs to replace $q_\theta$ with $\bar q_\theta$.}

\begin{table}[t]
    \centering
    \begin{tabular}{lccccc}
      \toprule
      Group & $|G|$ & $F_q$ & ESS & ESS Gain vs.\ $I$  & Cost vs.\ $I$ \\
      \midrule
      $I$               & 1   & -2.3168308(1)  & $0.066 \pm 0.013$ & $1.0$          & 1.0\\
      $T_2$             & 4   & -2.31686946(9) & $0.384 \pm 0.004$ & $6.1 \pm 1.2$  & 1.2\\
      $T_4$             & 16  & -2.31688522(8) & $0.542 \pm 0.003$ & $8.6 \pm 1.7$  & 1.7\\
      $M_2$             & 2   & -2.3168561(1)  & $0.182 \pm 0.031$ & $2.9 \pm 0.7$  & 1.1\\
      $C_4$             & 4   & -2.31687830(8) & $0.381 \pm 0.037$ & $6.0 \pm 1.3$  & 1.2\\
      $D_4$             & 8   & -2.31688866(7) & $0.470 \pm 0.053$ & $7.4 \pm 1.6$  & 1.3\\
      $T_4 \times C_4$  & 64  & -2.31690352(5) & $0.808 \pm 0.001$ & $12.8 \pm 2.6$ & 4.0\\
      \bottomrule
    \end{tabular}
    \caption{
      Symmetry comparison for $L=96$ and $\beta=1.12$,
      evaluated using $10^6$ samples.
    }
    \label{tab:l96-symmetries}
  \end{table}
  
Table~\ref{tab:l96-symmetries} shows the effect of symmetrization on the sampling. A model was trained at $L=96$ and $\beta = 1.12$, and then $10^6$ configurations were generated. $I$ is a model that has not been symmetrized (as considered in the previous sections). 

We first notice that symmetrization, even with very small groups, dramatically improves ESS; the gain ranges from a factor of 3 to 6 for group sizes of $2\le|G|\le 4$, at a cost increase of only 10-20\%. For example, applying a group of 4 lattice rotations $C_4$ increases the ESS from 0.066 to 0.38. In general, the larger the group is, the more gain in ESS it provides, but this comes at the cost of increasing the simulation expense – for a group with 64 elements, the ESS reached a value of 0.81, but the cost increased 4 times compared to no symmetrization.

\begin{figure}[t]
      \centering
      \includegraphics[width=0.6\linewidth]{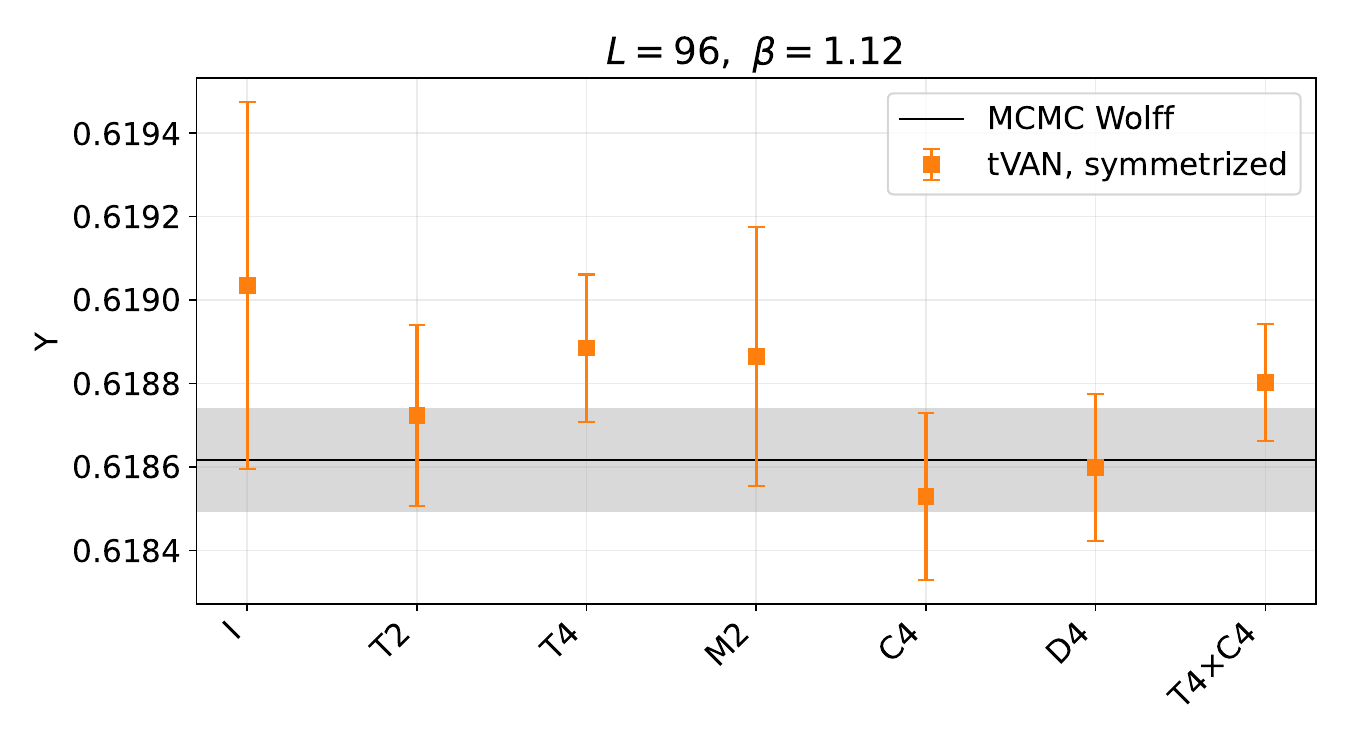}
      \caption{Helicity modulus $\Upsilon$ for $L=96$, $\beta = 1.12$ obtained from $10^6$ samples and different symmetrization groups.}
      \label{fig:helicity_through_symmetries}
  \end{figure}
  
In Figure~\ref{fig:helicity_through_symmetries}, we show the helicity modulus $\Upsilon$ obtained after  symmetrization with different groups. Parameters were chosen as above. All the results after symmetrization are consistent with the Wolff MCMC result. The reduction of uncertainties (obtained using the bootstrap method) is due to increased effective statistics and is consistent with the ESS values.

The results presented in this section were obtained with gauge fixing; however, we explicitly checked that without using it the gain from symmetrization is comparable. The reason for this is that lattice symmetries are already broken by the use of the conditional probability decomposition \eqref{eq:probability_density}; hence, the gauge fixing does not change this.

\subsection{tVAN as NF prior}

In section \ref{can_nf_tvan} we used NF with a simple prior distribution: uncorrelated variables drawn from a uniform distribution on a circle. This is the standard setup for NFs; however, it does not lead to a high-quality sampler.
Here we test an alternative approach, where, instead of a simple prior, we use the tVAN sampler. 
Since much of the system's structure is learned by tVAN, the NF only has to correct the remaining difference between the proposal and target distributions, instead of learning it from the ground up.
For a trivial NF, which is the identity transformation, we restore tVAN.

In Tab~\ref{tab:l32-tvan-nf} we show the comparison of a tVAN and an NF with a tVAN prior for $L=32$. Adding the NF increases the ESS only marginally. For larger system sizes, the difference is even smaller, or the NF even deteriorates the quality of sampling. We attribute this behavior to the low performance of the used NF architecture, as demonstrated in section \ref{can_nf_tvan}.

\begin{table}[h]
  \centering
  \small
  \begin{tabular}{lccccc}
    \toprule
    Model
      & $\mathrm{ESS}$\\
    \midrule
    tVAN
      & $0.826$\\
    tVAN$+$NF
      & $0.844$\\
    \bottomrule
  \end{tabular}
  \caption{Comparison at $L=32$ and $\beta=1.12$ using $10^6$ samples: pure tVAN vs. NF with a tVAN prior.}
  \label{tab:l32-tvan-nf}
\end{table}

\section{Summary}
\label{sect_summary}

In this work we have extended the transformer-based Variational Autoregressive
Network (tVAN) from discrete to continuous spin systems and applied it to the
two-dimensional XY model in the vicinity of the BKT transition. Each spin enters
the network as a pair $(\cos s_i,\sin s_i)$, and the conditional densities are parametrized by
mixtures of von Mises distributions. This choice provides normalized densities
on the circle that are easy to sample from and closely resemble the local
Boltzmann factor of the model, so that $K=4$ mixture components and a
single transformer layer already suffice.  The architecture was supplemented with
locality-informed attention, which restricts the attention span to the spins on
which the exact conditional distribution depends, and with $t^1$ approximate
probabilities, which for the XY model remain within the von Mises family and
can be combined with the network output analytically.

The resulting sampler outperforms existing neural samplers for the XY model by
a large margin. At $L=16$, where CAN and a normalizing flow reach an effective
sample size of order $10^{-2}$ or lower, tVAN reaches an ESS close to one, and at
$L=32$ its Metropolis--Hastings acceptance rate of about $80\%$ should be
compared with the $7.2\%$ reported for AdvNF. Approximate probabilities
substantially improve training, while fixing the global $O(2)$
rotation does not lead to a visible gain. We trained the model on lattices up to
$96\times 96$. The ESS decreases with the system size, from $0.98$ at $L=16$ to
$0.07$ at $L=96$, but the mode-dropping estimator remains consistent with one
for all studied sizes, showing that the decrease is not caused by missing parts
of the configuration space and can be compensated by NIS or NMCMC. Indeed, the
energy, magnetization and helicity modulus at $L=96$ agree with the results of
the Wolff algorithm within statistical uncertainties for all three studied
temperatures.

We have also shown that averaging the model probability over a group of
lattice symmetries after training provably does not decrease the ESS, and in
practice increases it dramatically. At $L=96$, averaging over four lattice
rotations raises the ESS from $0.07$ to $0.38$ at the cost of a $20\%$ increase
in the generation time, and averaging over 64 transformations yields an ESS of
$0.81$. Since the cost grows roughly linearly with the size of the group, small
groups offer the best gain per unit of computational time. As no training
is involved, this procedure can be applied to any autoregressive sampler at
negligible implementation cost. On the other hand, using tVAN as a prior for a
normalizing flow improves the ESS only marginally, which we attribute to the
limited expressivity of the flow architecture.

 A natural next step is the application of the method to lattice gauge theories. 
Given the  progress in NF-based samplers for lattice field theories, the combination of the tVAN prior with such architectures seems like a promising route.

\section*{Acknowledgments}
We gratefully acknowledge Polish high-performance computing infrastructure PLGrid (HPC Center: ACK Cyfronet AGH) for providing computer facilities and support within computational grant no. PLG/2025/018811. 

T.S. and D.Z. kindly acknowledge the support of the Polish National Science Center (NCN) Grant No.\,2021/43/D/ST2/03375.

\appendix

\section{$t^1$ approximate probabilities for the XY model }
\label{t1_priors_section}

The conditional density is written as
\begin{equation}
    p\left(s_i\mid s_{<i}\right) = \frac{ \displaystyle \int d s_{>i}\, e^{-\beta E(\bs)}}{\displaystyle \int ds_{\geq i}\, e^{-\beta E(\bs)}},
\end{equation}
where $ d s_{>i}=ds_{i+1}ds_{i+2}\ldots ds_N$.
Since the denominator does not depend on $s_i$,
\begin{equation}
    p\left(s_i\mid s_{<i}\right) \propto \int d s_{>i}\, e^{-\beta E(\bs)}.
    \label{ap_conditional_density}
\end{equation}
Decomposing the energy into an internal part $E\left(s_{<i}\right)$, which contains interactions only between already generated spins, and $E\left(s_{\geq i}\right)$, which contains the remaining contribution to $E$,  we get
\begin{equation}
    p\left(s_i\mid s_{<i}\right) \propto e^{-\beta E\left(s_{<i}\right)}\int d s_{>i}\, e^{-\beta E\left(s_{\geq i}\right)}.
\end{equation}
The factor of $E\left(s_{<i}\right)$ does not depend on $s_i$ and can be absorbed into a normalization constant,
\begin{equation}
    p\left(s_i\mid s_{<i}\right) \propto \int d s_{>i}\, e^{-\beta E\left(s_{\geq i}\right)}.
    \label{ap_conditional_density_integral}
\end{equation}
For the $t^1$ approximation, we consider interactions between $s_i$ and its already generated closest neighbors, denoted as $s_{i,\,left}$ and $s_{i,\,top}$ (see Figure  \ref{fig:p-cond-27}):
\begin{equation}
        E\left(s_i\right) = - \left( \cos(s_i - s_{i,\,left}) + \cos(s_i - s_{i,\, top}) \right).
\end{equation}
The $t^1$ approximation means $E\left(s_{\geq i}\right)\approx  E\left(s_i\right)$, substituting that into Eq.~\eqref{ap_conditional_density_integral}, we get
\begin{equation}
    p\left(s_i\mid s_{<i}\right)\approx  \tilde p\left(s_i\mid s_{<i}\right) \propto \int ds_{>i}\, e^{\beta\left(\cos(s_i - s_{i,\,left}) + cos(s_i - s_{i,\, top}) \right)}.
\end{equation}
Since the integration happens over the future spins $s_{>i}$, $\tilde p$ can be rewritten as
\begin{equation}
     \tilde p\left(s_i\mid s_{<i}\right) \propto e^{\beta\left(\cos(s_i - s_{i,\,left}) + \cos(s_i - s_{i,\, top}) \right)}. 
\label{ap_conditional_density_ap}
\end{equation}
This is because the integrand does not depend on the integration variables; hence, the integral is a constant that is absorbed in normalization. 

The sum of cosines could be transformed as follows
\begin{equation}
    \cos(s_i - s_{i,\,left}) + \cos(s_i - s_{i,\, top}) = A_i\cos(s_i) + B_i\sin(s_i)
\end{equation}
where
\begin{equation}
    A_i =  \cos(s_{i,\,left}) + \cos(s_{i,\,top}), \qquad B_i= \sin(s_{i,\,left}) + \sin(s_{i,\,top})
\end{equation}
Defining
\begin{equation}
    R_i = \sqrt{A_i^2 + B_i^2}, \qquad \mu_i^{AP} = \atant\left(B_i,A_i\right)
\end{equation}
yields
\begin{equation}
    A_i\cos(s_i) + B_i\sin(s_i) = R_i\cos\left(s_i - \mu_i^{AP}\right)
\end{equation}
By substituting the result into Eq.~\eqref{ap_conditional_density_ap}, we can see that the $t^1$ approximation is itself a von Mises distribution; with normalization, we have
\begin{equation}
    \widetilde p_i^{(1)}\left(s_i\mid s_{<i}\right) = \frac{e^{\kappa_i^{\mathrm{AP}}\cos(s_i-\mu_i^{\mathrm{AP}})}}
            {2\pi I_0(\kappa_i^{\mathrm{AP}})},
            \qquad \kappa_i^{\mathrm{AP}} = \beta R_i.
    \label{eq_p_tilde}
\end{equation}
Distribution \eqref{param_q_vonMises} is a mixture of von Mises distributions and, before applying $t^1$, has the form of 
\begin{equation}
    q_\theta^{(0)}\left(s_i\mid s_{<i}\right) = \sum_{j=1}^K \omega_{ij} \frac{e^{\kappa_{ij} \cos(s_i - \mu_{ij})}}{2\pi I_0\left(\kappa_{ij}\right)}.
\end{equation}
Multiplying $q_\theta^{(0)}$ by the $t^1$ factor \eqref{eq_p_tilde} yields us
\begin{equation}
    q_\theta\left(s_i\mid s_{<i}\right) \propto \sum_{j=1}^K \omega_{ij} \frac{e^{\kappa_{ij} \cos(s_i - \mu_{ij})}}{2\pi I_0\left(\kappa_{ij}\right)} \frac{e^{\kappa_i^{\mathrm{AP}}\cos\left(s_i-\mu_i^{\mathrm{AP}}\right)}}
            {2\pi I_0(\kappa_i^{\mathrm{AP}})}.
    \label{ap_prior_with_ap_not_simplified}
\end{equation}
By performing the same trigonometric operations on the exponents, we get
\begin{equation}
    \kappa_{ij}\cos(s_i - \mu_{ij}) + \kappa_i^{AP}\cos\left(s_i-\mu_i^{\mathrm{AP}}\right) = \tilde A_{ij}\cos(s_i) + \tilde B_{ij}\sin(s_i),
\end{equation}
where 
\begin{equation}
\begin{aligned}
    \tilde A_{ij} = \kappa_{ij}\cos(\mu_{ij}) +\kappa_{i}^{AP}\cos(\mu_i^{AP}), \\
    \tilde B_{ij} = \kappa_{ij}\sin(\mu_{ij}) +\kappa_{i}^{AP}\sin(\mu_i^{AP}).
\end{aligned}
\end{equation}
Defining
\begin{equation}
       \tilde \kappa_{ij} = \sqrt{\tilde A_{ij}^2 + \tilde B_{ij}^2}, \qquad \tilde \mu_{ij} = 
       \atant\left(\tilde B_{ij}, \tilde A_{ij}\right),
\end{equation}
we get
\begin{equation}
   \tilde A_{ij}\cos(s_i) + \tilde B_{ij}\sin(s_i) = \tilde \kappa_{ij}\cos(s_i - \tilde \mu_{ij}).
\end{equation}
Substituting into Eq.~\eqref{ap_prior_with_ap_not_simplified}, we get
\begin{equation}
    q_\theta\left(s_i\mid s_{<i}\right) 
    \propto
    \sum_{j=1}^K \omega_{ij}
    \frac{I_0(\tilde \kappa_{ij})}
    {2\pi I_0\left(\kappa_{ij}\right)I_0(\kappa_i^{\mathrm{AP}})} \frac{e^{\tilde \kappa_{ij}\cos(s_i-\tilde \mu_{ij})}}
    {2\pi I_0(\tilde \kappa_{ij})}
\end{equation}
Since $2\pi I_0(\kappa_i^{\mathrm{AP}})$ is shared across all of the mixture components, it can be absorbed into the normalizing constant
\begin{equation}
    q_\theta\left(s_i\mid s_{<i}\right) 
    \propto
    \sum_{j=1}^K \omega_{ij}
    \frac{I_0(\tilde \kappa_{ij})}
    {I_0\left(\kappa_{ij}\right)} 
     \operatorname{VM}\left(s_i; \, \tilde \mu_{ij}, \, \tilde \kappa_{ij}  \right)
\end{equation}
Mixture weights are then normalized, and we get
\begin{equation}
    \tilde \omega_{ij} = \frac{\displaystyle \omega_{ij} \frac{I_0(\tilde \kappa_{ij})}
    {I_0\left(\kappa_{ij}\right)} }{\displaystyle \sum_{l=1}^K \omega_{il}\frac{ I_0(\tilde \kappa_{il})}
    {I_0\left(\kappa_{il}\right)} }.
\end{equation}
Finally, the conditional density is expressed as \eqref{eq_q_decomp_AP}:
\begin{equation}
    q_\theta\left(s_i\mid s_{<i}\right) = \sum_{j=1}^K \tilde \omega_{ij}
    \operatorname{VM}\left(s_i; \, \tilde \mu_{ij}, \, \tilde \kappa_{ij}\right).
\end{equation}

\section{Proof that  $\textrm{ESS}_{\bar q} \geq \textrm{ESS}_q$}
\label{appendix_symmetriza}

We consider a group of transformations $G$ that do not change the energy: $E(g\b s) = E(\b s), \, g\in G$. It follows that the target probability $p$ is also invariant under the action of $G$. The distribution $\qt$, on the other hand, does not generally respect those symmetries. In section \ref{sec:symmetrization} we have introduced a symmetrized distribution
\begin{equation}
    \bar q_\theta(\bs) = \frac{1}{|G|} \sum_{g \in G} \qt(g \bs).
\end{equation}
By integrating both sides, we get
\begin{equation}
    \int \bar q_\theta(\bs) d \bs = \frac{1}{|G|} \sum_{g \in G} \int q_\theta(g \bs)d \bs
\end{equation}
Changing variables $gx = y$, we get 
\begin{equation}
    \int \bar q_\theta(\b s) d\b s = \frac{1}{|G|} \sum_{g \in G} \int q_\theta(y)dy 
    =
    \frac{1}{|G|}\sum_{g \in G} 1= 1
\end{equation}
meaning that $\bar q_\theta$ is a normalized probability density. In the derivation, we have used the fact that the integration measure is invariant under the group transformations. 

It is also easy to prove that $\bar q_\theta(\bs)$ is invariant under the action of the group $G$. By definition 
\begin{equation}
    \displaystyle \bar q_\theta(h \bs) = \frac{1}{|G|} \sum_{g \in G} q_\theta(h g \bs), \quad h\in G. 
\end{equation}
Since $g \mapsto h g$ is just a permutation of $G$, $\bar q_\theta(h \bs) = \bar q_\theta(\bs), \, h\in G$.
This means that the "symmetrized" weights
\begin{equation}
\bar w(\bs) = \displaystyle Z\frac{p(\bs)}{\bar q_\theta(\bs)}=\frac{e^{-\beta E(\mathbf{s})}}{\bar q_\theta(\mathbf{s})}
\end{equation}
are invariant under transformations in $G$.  

By definition
\begin{equation}
\label{eq:ess-q}
    \textrm{ESS}_{q_\theta} = \displaystyle\frac{\left\langle w\right\rangle_{q_\theta}^2}{\left\langle w^2\right\rangle_{q_\theta}}
\end{equation}
where unsymmetrized weights were defined in Eq.~\ref{eq. weight}.

We can write
\begin{equation}
\begin{aligned}
    \left\langle w\right\rangle_{q_\theta}^2 &= 
    \left( Z \int q_\theta(\bs) \frac{p(\bs)}{q_\theta(\bs)}\dl\bs \right)^2
       =
    \left( Z\int p(\bs)\dl\bs \right)^2
    = Z^2 \\
    \left\langle w^2\right\rangle_{q_\theta} &=  Z^2\int q_\theta(\bs) \frac{p^2(\bs)}{q_\theta^2(\bs)}\dl\bs 
    = Z^2\int \frac{p^2(\bs)}{q_\theta(\bs)}\dl\bs \equiv Z^2 I_q
\end{aligned}
\end{equation}
Substituting into \eqref{eq:ess-q}, we get
\begin{equation}
\textrm{ESS}_q = \frac{1}{I_q}\quad\text{and}\quad \textrm{ESS}_{\bar q} = \frac{1}{I_{\bar q}}
    \label{ess_integral_relation}
\end{equation}
where $I_{\bar q}$ has the same form as $I_q$ with $\bar q_\theta$ substituted in place of $q_\theta$.

For $y>0$, the function $\varphi(y)=\frac{1}{y}$ is a convex function, so Jensen's inequality
\begin{equation}
\varphi\left(\sum_i \lambda_i y_i\right)\leq \sum_i \lambda_i \varphi(y_i),\qquad \lambda_i\geq 0,\; \sum_i \lambda_i = 1
\end{equation}
translates to
\begin{equation}
\frac{1}{\sum_i \lambda_i y_i}\leq \sum_i \lambda_i \frac{1}{y_i}. 
\end{equation}
Substituting $y_i=q(\bs_i)$, we obtain
\begin{equation}
\frac{1}{\sum_i \lambda_i q(\bs_i)}\leq \sum_i \lambda_i \frac{1}{q(\bs_i)} 
\end{equation}
Now set 
\begin{equation}
\bs_g = g \bs,\qquad g\in G, \lambda_g =\frac{1}{|G|}    
\end{equation}
to obtain 
\begin{equation}
\int\dl \bs 
\frac{p^2(\bs)}{|G|^{-1}\sum_g q(g \bs)}\leq |G|^{-1} \sum_g \int \dl \bs  \frac{p^2(\bs)}{q(g \bs)} 
\end{equation}
as the multiplication by $p^2(\bs)>0$ and integration do not change the inequality. 
The left-hand side of this inequality is
\begin{equation}
\int\dl \bs 
\frac{p^2(\bs)}{\bar q( \bs)}=I_{\bar q}
\end{equation}
by definition. The right-hand side 
\begin{equation}
|G|^{-1}  \, 
 \sum_g \int \dl\bs  \frac{p^2(\bs)}{q(g \bs)} = |G|^{-1} \sum_g \int \dl \bs  \frac{p^2( g \bs)}{q(g \bs)} = 
|G|^{-1} \sum_g \int \dl\bs  \frac{p^2(\bs)}{q(\bs)}=|G|^{-1} \sum_g I_q =I_q,
\end{equation}
where we have used the fact that $p(g\bs)=p( \bs)$ and the invariance of the measure under the group transformations. 
This proves that $I_q \ge I_{\bar q}$, hence 
\begin{equation}
\textrm{ESS}_q \leq \textrm{ESS}_{\bar q}.     
\end{equation}

\section{Hyperparameters}
\label{appendix_hyperparameters}

\begin{figure}[t]
    \centering
    \includegraphics[width=0.49\linewidth]{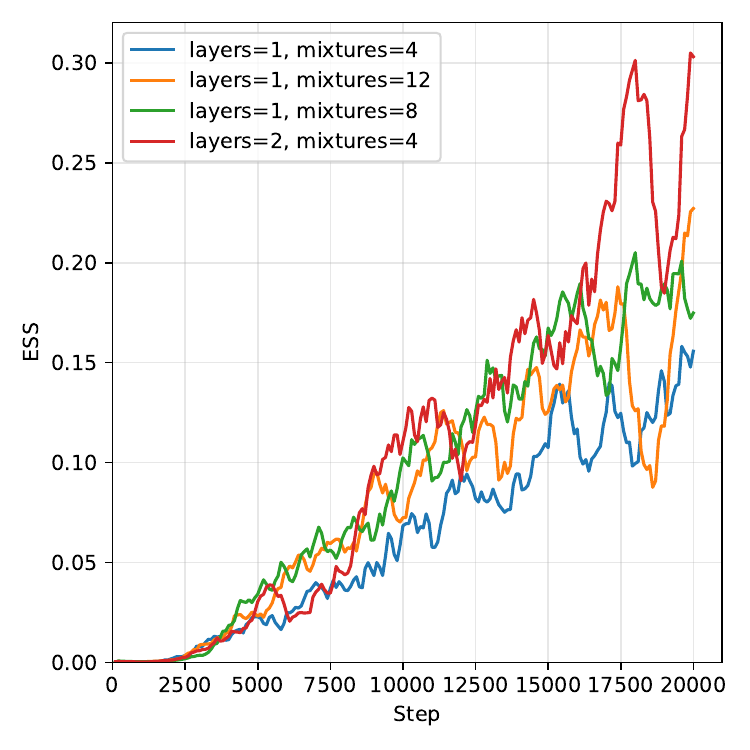}
    \includegraphics[width=0.49\linewidth]{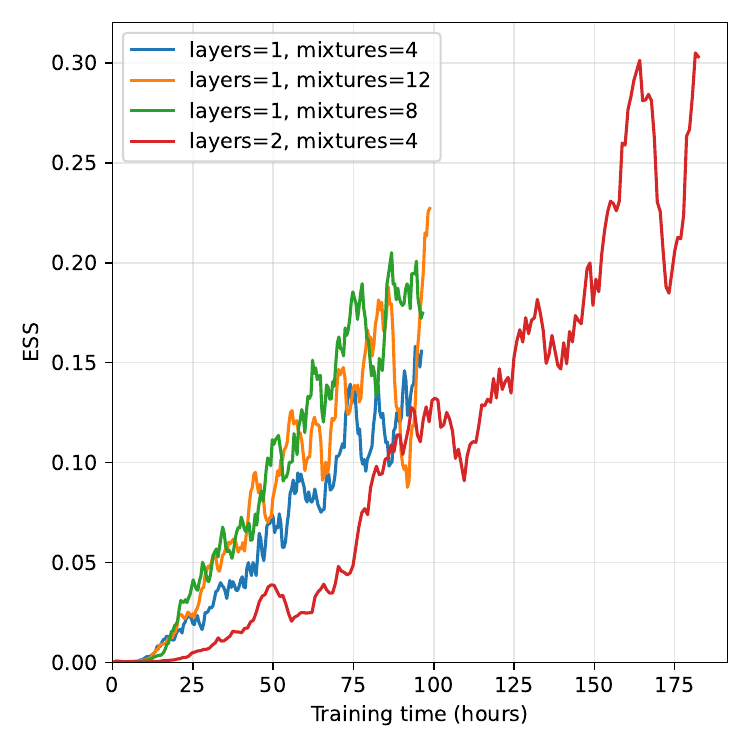}
    
    \caption{ESS comparison of single-layer models with $K =4, 8, 12$ mixture components and a two-layer model with $K=4$ mixture size  across steps (left panel) and training time (right panel).}
    \label{fig:hyperparams_ess_step_mean_1k}
\end{figure}

The main text works with results obtained on a simple model composed of a single transformer layer and $K=4$ components for the von Mises mixture. To examine the dependence on model expressivity, we additionally performed tests with different number of mixture components, namely $K= 8, 12$, and two transformer layers, while the remaining setup was left unchanged. 

We analyze the model's ESS through training steps as well as training time (see Fig~\ref{fig:hyperparams_ess_step_mean_1k}). 
The runs were performed at $L=96$ with inverse temperature $\beta = 1.12$ on a single A100 (40GB) GPU. We do not observe statistically significant differences between training for different $K$. The 2-layer transformer reaches a higher ESS when trained for 20k steps, but at the expense of a longer training time. We therefore conclude that $K=4$ and one transformer layer is a reasonable choice.

\bibliography{references2.bib}

\end{document}